\documentclass[aps,twocolumn,showpacs,superscriptaddress,a4paper]{revtex4-1}

\usepackage[dvips]{graphicx}
\usepackage{amsmath}
\usepackage{amssymb}

\usepackage{hyperref}

\usepackage{geometry}
\begin{document}
\bibliographystyle{apsrev}

\title{Order-by-disorder in classical oscillator systems}

\begin{abstract}
We consider classical nonlinear oscillators on hexagonal
lattices. When the coupling  between the elements is repulsive, we observe
coexisting states, each one with its own basin of attraction. These states
differ by their degree of synchronization and by patterns of phase-locked
motion. When disorder is introduced into the system by additive or
multiplicative Gaussian noise, we observe a non-monotonic dependence of the degree
of order in the system as a function of the noise intensity: intervals of noise intensity with
low synchronization between the oscillators alternate with intervals where more
oscillators are synchronized. In the latter case, noise induces a higher degree
of order in the sense of a larger number of nearly coinciding phases. This
order-by-disorder effect is reminiscent to the analogous phenomenon known from
spin systems. Surprisingly, this non-monotonic evolution of the degree of order
is found not only for a single interval of intermediate noise strength, but
repeatedly as a function of increasing noise intensity. We observe noise-driven
migration of oscillator phases in a rough potential landscape.
\end{abstract}

\author{F. Ionita}
\affiliation{School of Engineering and Science, Jacobs University Bremen, 28759 Bremen, Germany}

\author{D. Labavi\'{c}}
\affiliation{School of Engineering and Science, Jacobs University Bremen, 28759 Bremen, Germany}

\author{M. A. Zaks}
\affiliation{Institut f\"ur Mathematik, Humboldt University Berlin, 12489 Berlin Germany}

\author{H. Meyer-Ortmanns}
\affiliation{School of Engineering and Science, Jacobs University Bremen, 28759 Bremen, Germany}


\maketitle
%

\section{Introduction}
\label{sec:intro}
The so-called order-by-disorder effect is usually
discussed in different contexts of spin models.
The notion refers to the frequently found situation
in which the ground state is degenerate due to the competition
among the interactions, and this degeneracy is lifted  due to
disorder. Here the lifting can be temperature driven (that is entropically)
\cite{26,29} or quantum driven as in \cite{barnett} or \cite{reimers,chubukov}.
The term itself was introduced in classical spin models \cite{26},
and was further discussed in \cite{27}; it has been also observed
in quantum magnetism \cite{29,28} and
in ultracold atoms, see for example \cite{30}.
In  \cite{reimers,chubukov} this effect was studied for a two-dimensional
Heisenberg antiferromagnet on a Kagom{\'e} lattice, where the long-range order
of spins was induced via disorder. As mentioned in \cite{29}, it depends
crucially on the degree of degeneracy whether the order-by-disorder effect
dominates in the competition of interactions so that its implications on
correlations become visible. What makes  the order-by-disorder effect
particularly interesting is the feature that it appears to be counterintuitive.

In this paper we consider a different realization of an order-by-disorder effect
in a multistable system of classical active rotators \cite{15}.
Active rotators are a prototype of systems that,
depending on the choice of parameters, display
either periodic oscillations or excitable fixed-point behavior.
Most studies of ensembles of rotators treated attractive (positive) coupling,
which favored in-phase synchronization.
In a few studies a random subset of attractive
couplings was replaced by repulsive ones. Disorder in the very sign of couplings
was the focus of \cite{183} with the result that a moderate fraction of
repulsive interactions triggered global firing of the rotator ensemble,
an effect that is suppressed for large networks
unless the interaction topology is appropriately changed.

In previous studies the repulsive coupling was introduced into the ensembles of
active rotators \cite{183} and Kuramoto oscillators \cite{zanette} at random,
without an explicit control over the induced frustration. In a recent work
\cite{hong} the case of directed links was treated: for a certain proportion of
oscillators (``contrarians'') the coupling to the mean field was negative, which
resulted in rich nontrivial dynamics. Differently, in this paper we consider
undirected coupling and focus on effects that are merely induced by
frustration, while disorder is implemented in the form of additive or multiplicative noise.
We therefore prescribe the sign of couplings in a way that it is possible
to control the number of induced frustrated bonds
and to distinguish effects, generated by frustration and noise,
from those which owe to disorder in the coupling signs
in combination with noise.
As we shall see, for the latter case the order-by-disorder effect is absent.

The notion of frustration in oscillatory systems was introduced
in \cite{daido} and further used in \cite{zanette} for Kuramoto oscillators
with undirected coupling. Later it was generalized by one of us \cite{pablo}
for directed networks and excitable and oscillatory systems.
In \cite{pablo} we have shown that frustration in these systems can indeed
lead to a considerable increase in the number of stationary states
and to multistability,
an effect analogous to that in spin systems. It is this growth
in the number of coexisting attractors,
for which we here introduce frustration in the active rotator systems.

Without noise the system obeys a gradient dynamics which drives it
to lower values of the potential in the high-dimensional configuration space.
The evolution either ends up in one of the local minima of the energy landscape,
or (since the potential is not bounded from below) proceeds {\it ad infinitum}
along the descending rifts at the bottom of landscape valleys.
While the local minima in the energy landscapes of spin systems correspond
to different fixed points, attractors for active rotator units,
depending on the choice of parameters, can be either fixed points
(minima as well) or trajectories along the rifts.
If, in the toroidal phase space of the system, the rifts turn into
closed curves, the respective attractors are limit cycles:
the motion along them corresponds to periodic oscillations.
Non-closed rifts correspond to temporal patterns which never repeat
themselves exactly: either quasiperiodic or chaotic oscillations.
Besides the quantitative details, the attractors differ by their degree
of synchronization
(which we call ``order"): the number of  {\it different}
oscillator phases in the patterns of phase-locked motions.
The lower this number is, the higher the order,
the more phases in the ensemble coincide in a state.
Counterintuitive, this order can be increased
via the introduction of disorder into the system:
disorder in the form of additive or multiplicative white noise.

\section{The model}
\label{sec:model}
\begin{figure}
\centering
\includegraphics[width=80mm]{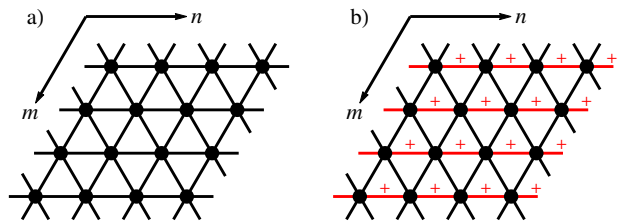}
\caption{Hexagonal lattice with all triangles frustrated for all couplings being negative (a) and not frustrated (b) for positive couplings along the horizontal links and negative ones otherwise.}
\label{fig.1}
\end{figure}

We study systems of $N$ active rotators~\cite{15},
whose phases $\varphi_i$ are governed by the equations:
\begin{eqnarray}
\label{eq_model}
\frac{d\varphi_i}{dt} &=&
\omega_i-b\sin{\varphi_i} + \sigma_A \xi_i(t) + \nonumber \\ & &
\frac{(\kappa+\sigma_M \eta_i(t))}{\mathcal{N}_i}\sum_j
A_{ij}\sin(\varphi_j-\varphi_i).
\end{eqnarray}
Here $\omega_i$ denote the natural ``frequencies'' of the rotators,
$b$ and $\kappa$ parameterize, respectively, the level of excitability
and the coupling strength. Further, $\mathcal{N}_i$ denotes the number of
neighbors to which the $i$-th unit is connected, and $A_{ij}$
is the adjacency matrix with $A_{ii}=0$, $A_{ij}=1$ if
$i\neq j$ and units $i$ and $j$ are connected, otherwise $A_{ij}=0$.
The terms $\sigma_M\eta_i$ and $\sigma_A\xi_i$ denote
multiplicative and additive white noise, respectively,
with $\langle\eta_i\rangle=0=\langle\xi_i\rangle$,
$\langle\eta_i(t)\eta_j(t)\rangle=2\delta_{ij}\delta(t-t^\prime)$,
likewise $\langle\xi_i\xi_j\rangle=2\delta_{ij}\delta(t-t^\prime)$,
where $\sigma_M$ and $\sigma_A$ are measures for the multiplicative and
additive noise intensities, respectively.
Throughout the paper we set $\omega_i=\omega$:
for a set of identical elements the synchronized solution always
exists on arbitrary plane lattices, in contrast to two-dimensional
lattices of oscillators with distributed frequencies~\cite{Hong_Park}.
We choose $A_{ij}$ to represent a hexagonal lattice of the size
$M\times L$ with periodic
boundary conditions, unless otherwise stated,
as shown in Fig.~\ref{fig.1}~(a),
so that each unit has the same number of nearest neighbors:
$\mathcal{N}_i=6$.

Most former studies have been performed for positive coupling
$\kappa=\vert\kappa\vert$, while we are mainly interested in repulsive coupling
and choose $\kappa=-\vert\kappa\vert$.
The reason is that negative $\kappa$ in combination with the hexagonal
coupling pattern turns all bonds into frustrated.
Consider any elementary triangle formed by units $i,j,k$.
Negative coupling ensures that unit $i$ prefers to be antiphase
to $j$ and $j$ prefers to be antiphase to $k$.
As a result, $i$ would be in phase to $k$, which is in conflict
with the repelling direct link between them.
So whatever phase the unit $k$ assumes,
the bond either between $k$ and $j$ or between $k$ and $i$ is frustrated.

In order to compare our results  with couplings that are disordered with
respect to their sign, but do not lead to any frustrated bonds, we show also the
case of positive couplings for the bonds along the horizontal directions in
Fig.~\ref{fig.1}~(b), while all other bonds are repulsive.

\section{Multistability for repulsive coupling}
\label{sec:multistability}
{\bf Kuramoto oscillators.}
Consider eq.~(\ref{eq_model}) with both sources of noise set to zero. Let us first
discuss the phase space structure for $b=0$: a set of $N$ Kuramoto oscillators
coupled on a hexagonal $(M\times L)$-lattice. By going into the comoving frame
via $\varphi(t)\rightarrow \varphi(t)-\omega t$ and rescaling the time via the
normalized coupling strength $|\kappa|/6$ (for the given number of nearest
neighbors), we arrive at
\begin{equation}
\label{eq_kuramoto}
\dot{\varphi}_i\;=\;{\rm sign}(\kappa)\,\sum_j A_{ij}\sin(\varphi_j-\varphi_i).
\end{equation}
While for positive $\kappa$ this system has a single stable fixed point
in which all phases are the same (this corresponds to synchronous oscillations
with frequency $\omega$ in the lab frame), we see hints on the expected
multistability for the negative coupling for a subset of solutions. These are
plane-wave solutions, characterized  by fronts of constant phases
along parallel lines on the hexagonal lattice
such that no nearest neighbors share the same phase.
Their spatial distribution is characterized by
\begin{equation}
\label{eq_plane_wave}
\varphi_{m,n}\;=\;\frac{2\pi}{M}k_1m+\frac{2\pi}{L}k_2n
\end{equation}
for coordinates $m=1,...,M$ and $n=1,...,L$,
where allowed values for the wave vector $k_1,k_2$ are restricted
to integers by the periodic boundary conditions.
In patterns of this kind the coupling term identically vanishes.
For this set we show in the Appendix A.1  that for a sufficiently large
extension of the lattice and an even number of the linear extension $M=L$,
there are always two sets of wave vectors $k_1=k_2=k$ and $k_1=k_2=k+1$
such that the plane waves correspond to different solutions.
These solutions do not merely differ by a rotation of the linear front,
but by the very number of oscillators sharing the same phase
(below we denote such sets of oscillators as ``clusters'').

Plane-wave solutions, which obviously reflect the lattice symmetry
in their fronts of constant phases, are not the only stable solutions on a hexagonal
lattice, as we shall discuss in more detail below.

{\bf Active rotators.}
In the following, without restriction of generality, we assume $b$
to be non-negative. For $b\neq 0$, by rescaling the time unit we set
$b=1$ and follow the solutions in the parameter space of  $\omega$ and $\kappa$.
For $\omega < 1$ and positive $\kappa$, active rotators have a stable fixed
point, $\varphi_i = \varphi_s=\arcsin{\omega}$, $i = 1,\ldots, N$, in which all
elements are synchronized with the same phase. This fixed point is stable for
sufficiently small negative $\kappa$ as well;
when $\kappa$ is decreased, it loses stability via the degenerate
pitchfork bifurcation at
$\kappa_c=-\sqrt{1 - \omega^2}/(1 - \lambda_{\text{min}}/\mathcal{N})$,
where $\lambda_{\text{min}}$ is the minimal (most negative)
eigenvalue of the adjacency matrix $A_{ij}$. The corresponding derivation
is provided in Appendix A.2. For the hexagonal
lattice the minimal eigenvalue is always degenerate; its multiplicity
depends on the periods $M$ and $L$: for $M\neq L$ and $M$=$L$=$3n$ it equals 2;
for $M$=$L\neq 3n$ it equals 6 with the exception of the case $M$=$L$=4
when the multiplicity of the minimal eigenvalue is 9. Accordingly, the dynamics
on the corresponding central manifold is not quite trivial. Leaving the
complete description for the forthcoming publication,
here we note only that, in spite of the high multiplicity,
the gradient character of the dynamics forbids the complexification of eigenvalues,
therefore no Hopf bifurcations can occur.

\begin{figure}
\includegraphics[width=80mm]{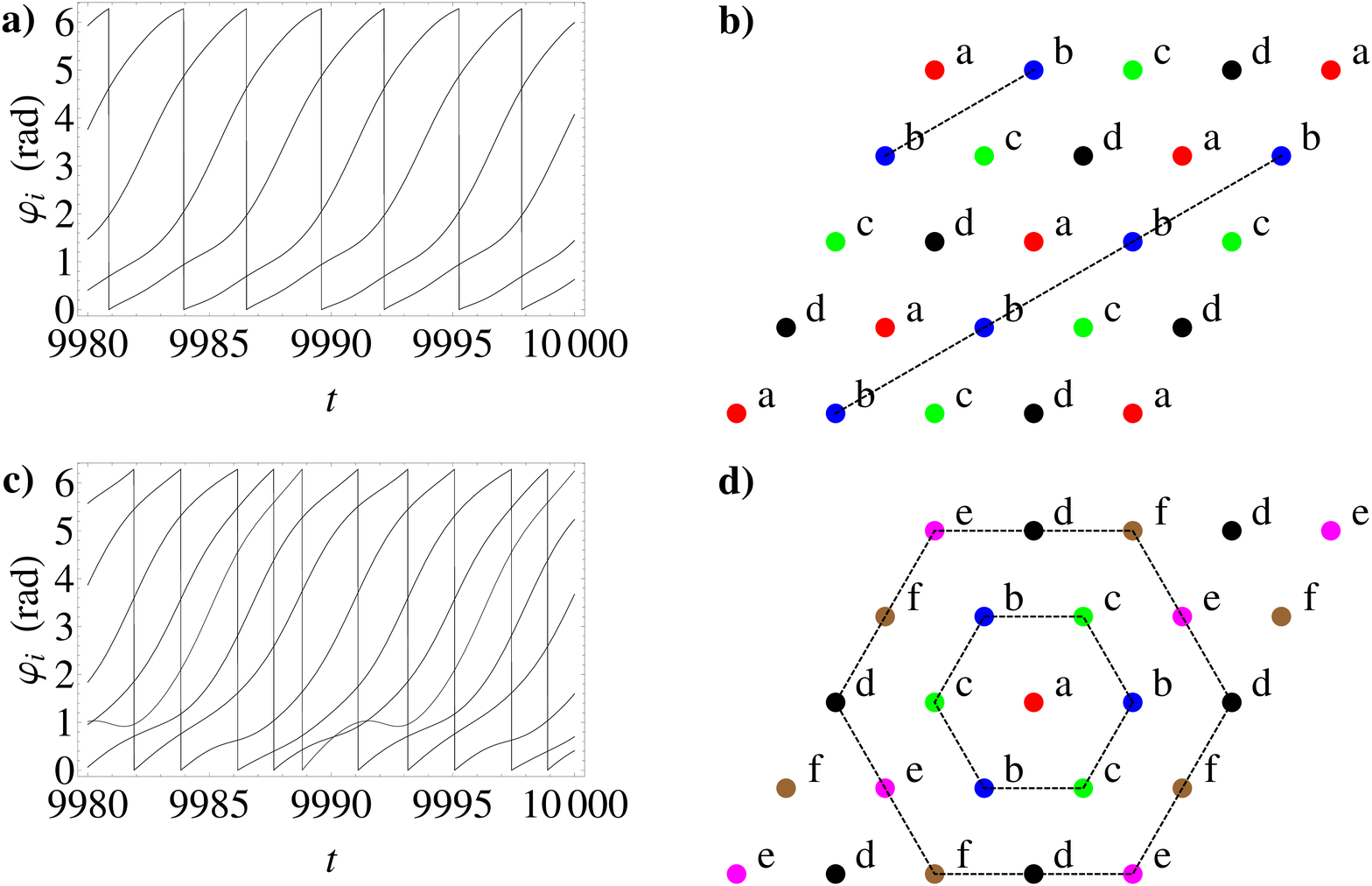}
\caption{Solutions of
eq.~(\ref{eq_model}) with  4 and 6 clusters on a $4 \times 4$ lattice
for $\omega =0.7$, $b=1$, $\kappa = -2$ and $\sigma_M=\sigma_A=0$.}
\label{fig.2}
\end{figure}

For sufficiently negative $\kappa$, when, at $\omega>1$, every individual unit
is in the limit-cycle state, there are no stable fixed points. Since we are
primarily interested in the order-by-disorder effect, we do not further zoom
into the bifurcation region, but choose $\kappa$ sufficiently negative as to be
well inside the parameter domain with time-dependent dynamics. This domain is
characterized by coexisting states with different synchronization patterns which
perform periodic or quasiperiodic oscillations.
Even for small lattice sizes the number of coexisting
attractors can be quite large, and below we
briefly describe just two typical patterns.

Similarly as for Kuramoto oscillators, one form of
stable solutions are {\bf plane-wave patterns}.
They are characterized by fronts of identical phases
which are parallel to each other and extend along straight lines
of the hexagonal lattice as shown in Fig.~\ref{fig.2}~(b), connecting sites
which are not nearest neighbors in accordance with the repulsive couplings.
A special case of plane wave solutions are splay-like states for which
the instantaneous values of phases characterizing different clusters
are separated on the unit circle by equal intervals.

Another type of solution appears as a {\bf spherical wave} on a
hexagonal lattice. For example, it is found as a 6-cluster solution on a
$4n_1\times4n_2$ lattice with integer $n_1,n_2$.
At a first glance, the phase assignments on the
$4\times 4$-lattice, displayed in Fig.~\ref{fig.2}~(d),
may look irregular;
there the characters $a,...,f$ label different clusters.
However, what we see is a
sector of a ``discretized" spherical wave on the hexagonal lattice, with an
isolated oscillator in its center, surrounded by two clusters of three
oscillators, in alternating order $bc\;bc\;bc$, along the first polygon ring of
nearest neighbors. In the second ring of next-nearest-neighbors of the center,
three clusters of three oscillators are arranged like $edfedfedf$.
Phase differences within a ring and between different rings do not stay constant,
but oscillate regularly.
Placing on a $4\times 4$-lattice the center of the wave into
any of the 16 sites and performing any number of 60$^\circ$ rotations,
we obtain 96 replicas of this configuration.

The existence of the described wave
assumes compatibility with the lattice size and the boundary conditions;
its stability depends on the strength of repulsive coupling.

{\bf Characterization of states.} In view of our goal to analyze the
order-by-disorder effect, it is sufficient to distinguish the states
according to their cluster partitions: we denote the states as $p_n$,
$n$ being the overall number of clusters.
We say that several oscillators belong to the same
cluster if they share the same phase within numerical accuracy.
This accuracy will be lowered and  adapted in the presence of noise.
Of course, this characterization is not unique: several different states
can have (and as we will see below they {\em do} have indeed) the same
partition into clusters. By {\em different} we mean dynamical distinctions:
two states which cannot be transformed into each other by a combination of
symmetry transformations of the lattice (translations, rotations, reflections);
for periodic solutions, it suffices to have non-coinciding values of the period.


\section{Order-by-disorder for active rotators}
\label{sec:order-by-disorder}
{\bf Quasi-stationary states.}
In the following we consider active rotators coupled with frustration on
hexagonal lattices under the action of noise. Without noise, the stationary
$p_n$-states, characterized by their cluster partitions, correspond to
frequency - synchronized oscillators, with phase-locked motion in case of Kuramoto
oscillators, (that is time-constant phase differences between different
clusters,) and oscillating phase differences between different
clusters in case of active rotators.

In contrast, under the action of weak and moderate
noise, the system exhibits behavior which we would call quasi-stationary.
In fact, this is noisy dynamics on a landscape with many traps:
the trajectory moves from one $p_n$-state to another, spends there a
noticeable amount of time, departs to the next one, and this ad
infinitum. For strong noise the short intervals of stay can hardly be resolved
and the motion turns into a random walk across the rough energy landscape.

We represent our results in terms of time evolutions of individual
oscillator phases $\varphi_i(t), i=1,...,N$, a presentation
which is limited by the spatial resolution for larger systems.
The widespread characterization in terms of the
Kuramoto order parameters $\rho_n=1/N\sum_{j=1}^N \exp{i n \varphi_j}$
has been designed for the case of uniformly rotating oscillators,
and works especially well in the splay situation, when
the phase clusters are more or less uniformly
placed on the circle. There for the state $p_n$
the values of $|\rho_j|$ with $j<n$ (nearly) vanish, whereas
the value of $|\rho_n|$ is close to 1. Since in our computations
the values of phases in the clusters are typically not equally spread,
the lower parameters $|\rho_j|$ are usually not small
and can hardly serve as simple indicators.
In case of active rotators
the set of order parameters is even less appropriate: phase space of an
active rotator involves segments of relatively fast and relatively slow
evolution. Since major portions of time are spent on slow segments,
the parameter $\rho_1$ would be close to 1 (and, hence, indicate synchrony)
even in the case when the coupling is completely absent. It may
also happen that the average over one time period of some
$\rho_n, n\in\{1,...,N\}$ takes a larger value for a state with less
coinciding phases. Examples will be discussed below.

Our basic example is the lattice of size $4\times 4$ at $\omega$=0.7,
$b$=1 and sufficiently strong negative coupling $\kappa$=--2.
In the absence of noise, this lattice is especially multistable.
Starting from different initial conditions (a
sequence of randomly generated 2$\times 10^4$ sets) we were able to resolve here
75 stable periodic orbits of the type $p_4$ which have different
values of the period and, hence, are dynamically different.
Each of these orbits has a multitude of symmetric replicas.
Besides, we detected two dynamically different spherical waves of the type $p_6$:
one of them corresponds to a limit cycle, and another one --
to a quasiperiodic state.
Finally, one periodic solution and several
quasiperiodic states belong to the type $p_{16}$:
all instantaneous values of individual phases are different.
The reason why we call the latter type a
clustered pattern although each cluster consists of a single oscillator,
is the fact that in this state the phases are still correlated:
phase differences between the sixteen oscillators oscillate periodically
or quasiperiodically. This regular character of oscillations
persists at weak noise. In contrast,  in the most disordered state that is
observed for strong noise, the evolution of individual phases seems fully
uncorrelated.

For detection of periodic and quasiperiodic states we used the numerically
obtained Poincar\'e mapping on the hypersurface $\varphi_1$=const.
Limit cycles turn into attracting fixed points of the mapping,
whereas the quasiperiodic states (two-dimensional tori) are
identified as smooth curves on the secant surface.

\begin{figure*}
\centering
\includegraphics[width=160mm]{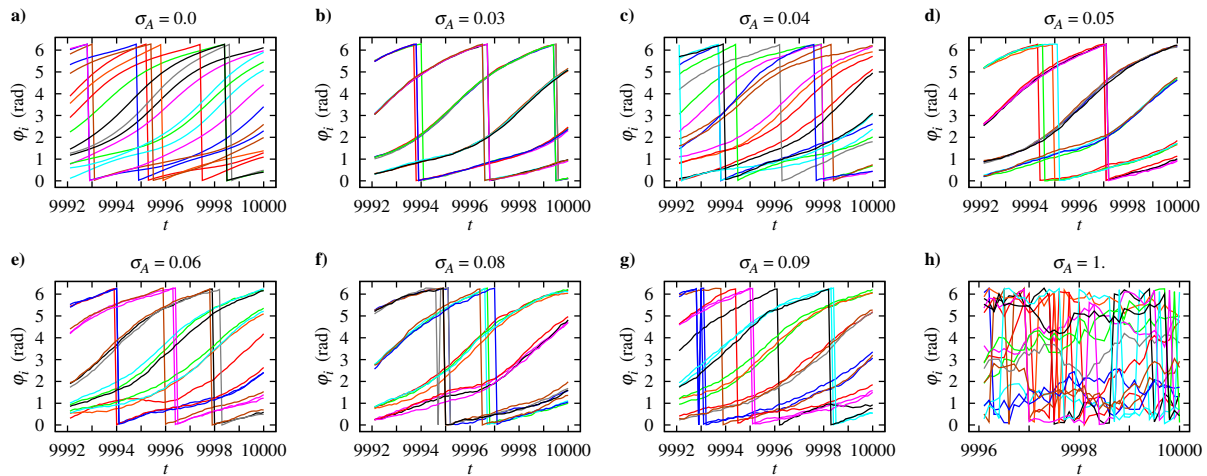}
\caption{Order-by-disorder on a $4 \times 4 $ lattice of active rotators, for $\omega = 0.7$,
$b=1$, $\kappa = -2$, $\sigma_M$=0 and monotonically increasing noise intensity
$\sigma_A$ between panels (a) to (h). For further explanations see the text. }
\label{fig.3}
\end{figure*}

Now we turn on additive noise and monotonically increase its intensity.
The results are shown in Fig.~\ref{fig.3}.

We plot the phases  of all $16$ oscillators in the interval $[0,2\pi]$ as
functions of time. While the noise intensity is varied from $\sigma_A = 0.01$ to
$\sigma_A = 1$, all other parameters are kept fixed: $\omega=0.7$, $b=1$,
$\kappa=-2$. In the presence of noise, exact coincidence of phases is washed
out, but the tendency to form easily recognizable groups persists, therefore we
soften the definition: we say that oscillators belong to one cluster if their
phases agree within $1.6 \cdot 10^{-1}$ for $\sigma_A=0.02$  and within
$5.5\cdot 10^{-1}$ for $\sigma_A=0.1$, so that the accuracy has to be adapted to
the noise intensity. If we characterize the solutions by the cluster partitions
in terms of $p_n$,  and mark all disordered states  by the symbol $d$, we read
off the following sequence from Fig.~\ref{fig.3}: $p_{16}$ (each oscillator is
isolated) for $0 \leq \sigma\leq 0.02$; $p_4$ (four clusters with four
oscillators in each), for $\sigma_A=0.03$, $\sigma_A=0.05$, and later again for
$\sigma_A=0.08$; $p_6$ solutions consisting of one isolated oscillator, two
clusters with three oscillators each and three clusters with three oscillators
each around the center of the spherical wave, for $\sigma_A=0.06$,
($\sigma_A=0.07$, not displayed), $\sigma_A=0.09$ (and $\sigma_A=0.1$, not
displayed). The states as seen for $\sigma_A=0.03$ and $\sigma_A=0.05$
correspond to $p_4$ patterns, but for stronger noise the system stays in a more
ordered realization of this pattern, which may give a hint on a deeper valley of
the attractor in which the synchronization is less sensitive to the noise. For
even stronger noise like $\sigma_A=1.0$ the solutions get fully disordered: all
phases are non-synchronized and uncorrelated. In terms of our notion of order, a
4-cluster solution is more ordered than a 6-cluster solution which is more
ordered than a 16-cluster solution.

{\bf Generic features.} For all realizations, displayed in Fig.~\ref{fig.3}, we
varied only the intensity of additive noise, starting from the same randomly
chosen initial condition, and choosing the same seed for the random number
generator. The question therefore arises of how representative are these plots.
Individual plots like those of Fig.~\ref{fig.3} do depend on the boundary
conditions (whether periodic or not), initial conditions (random or not), the
realization of noise (additive, multiplicative, weak, intermediate, and the very
realization), moreover on the time instant at which the snapshots are taken, and
the lattice size. What is independent of the concrete choice of all these
conditions is the observed non-monotonically varying degree of order when the
noise intensity is monotonically increased. This feature should be observed as
long as the boundary and initial conditions allow for many coexisting attractors
in the system without noise, as we shall explain in the following.

We formed an ensemble of hundred randomly chosen initial conditions, and for
each of them we repeated the simulation with ten different realizations of noise
of the same intensity. Let us call the sequence
$(p_{16}$, $p_4$, $p_{16}$, $p_4$, $p_6$, $p_4$, $p_6$, $p_6$, $p_4$, $p_6$, $p_6$, $d)$
that is observed in Fig.~\ref{fig.3} a ``pattern".
If we view as ``qualitatively the same pattern" those sequences
which differ by permutations of the sequence or a different number of 6-cluster
and 4-cluster solutions, then out of the hundred initial conditions  $\sim 20\%$
led to qualitatively the same pattern as in Fig.~\ref{fig.3}. The remaining $80\%$ of
initial conditions lead to sequences like $(p_{16},\ldots,p_6,\ldots,d)$ with
homogeneous behavior for a larger intermediate range of noise, in case of which
the non-monotonic behavior is less pronounced. From the different realizations
of noise for otherwise unchanged parameters about $30\%$ have led to
qualitatively  the same patterns, the actual numbers vary between one and six
out of ten. We conclude that the observation of a sequence with a pattern as displayed in
Fig.~\ref{fig.3} is not a rare event in the stochastic dynamics of the
ensemble.

It should be noticed that the patterns of Fig.~\ref{fig.3} correspond to
a short time window after $10^4$ time units  from the
start of integration and are not meant as final states.
The sequence of order may differ for other, earlier and also (arbitrarily) later
time windows, as we explain below. In particular, on sufficiently long time segments
the $p_4$ solution of panel b) will repeatedly transform to $p_6$ and $p_{16}$ solution
and back.

\begin{figure*}
\begin{center}
\includegraphics[width=140mm]{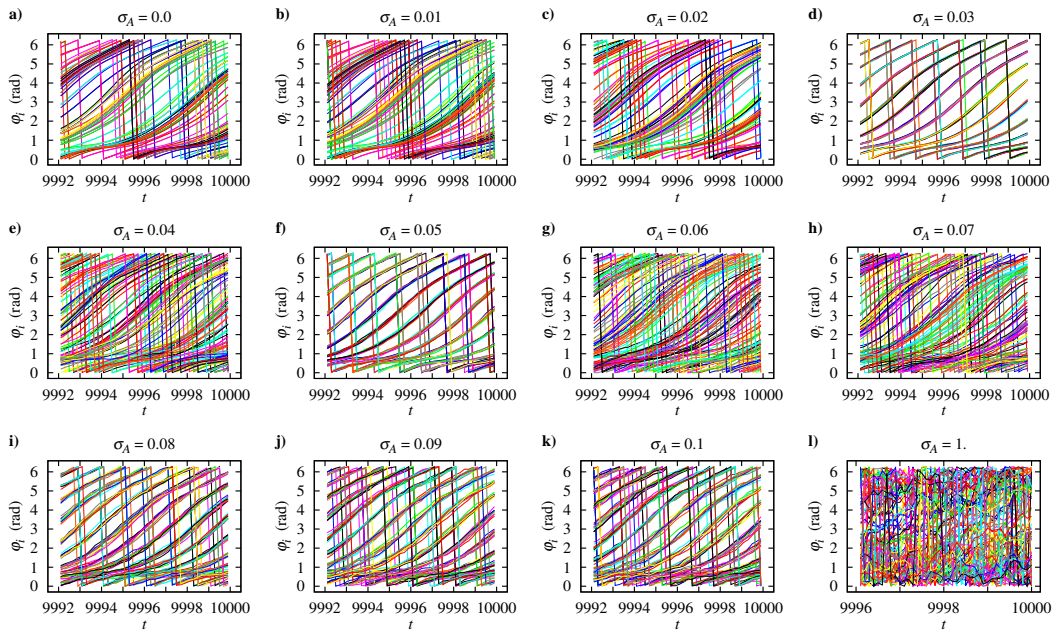}
\end{center}
\caption{Order-by-disorder on a $10 \times 10$ lattice of active rotators, for parameter values
$\omega = 0.7$,  $b=1$, and $\kappa = -2$. }
\label{fig.4}
\end{figure*}

Similar, but visually more pronounced differences between more and less ordered
and disordered states are recovered on a $10\times10$-lattice. Ordered states
with ten clusters of ten oscillators each are seen for the intermediate noise
intensities $\sigma=0.03,0.05,0.08,0.09$ and $0.1$, while we see hundred
different phases, synchronized for zero or small noise values and fully
disordered for large noise values, for the plots see  (Fig.~\ref{fig.4}).

\begin{figure*}
\begin{center}
\includegraphics[width=110mm]{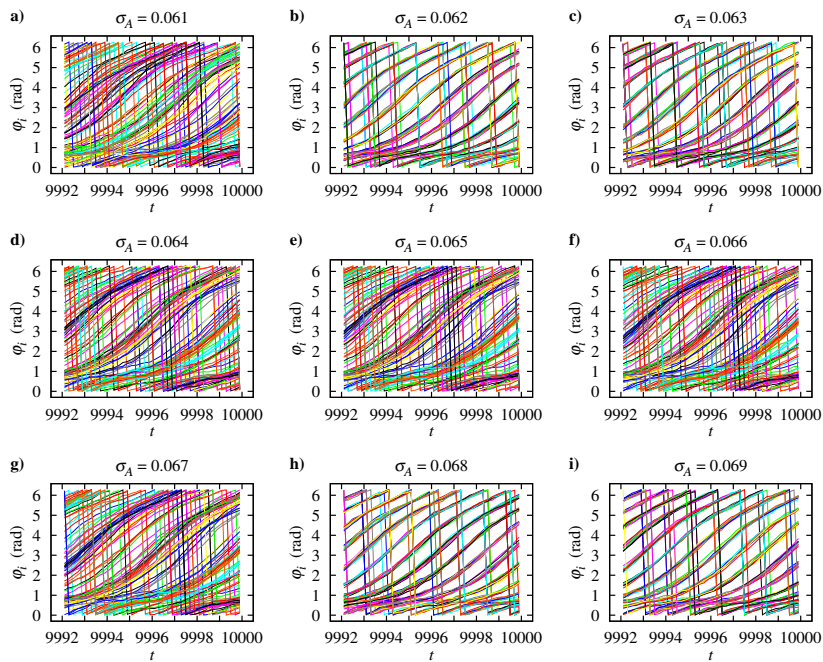}
\end{center}

\caption{Zoom into the noise interval between $0.061$ to $0.069$, for otherwise
the same parameters as in Fig.~\ref{fig.4}. }
\label{fig.5}
\end{figure*}

Remarkably, if we here further zoom into the noise intervals, for example into
$[0.06,0.07]$ between the disordered patterns of hundred different phases at
$0.06$ and $0.07$ (Fig.~\ref{fig.5}), we see a sequence like $d\;o\;o\;d\;d\;d\;d\;o\;o$, for
equidistant values of $\sigma_A$ between $0.061$ to $0.069$, where $d$ stands
for hundred different phases and $o$ (order) for ten different phases organized
in ten clusters.

At this point one may be tempted to continue the zooms to find out,
down to which resolution of the noise intensity  the characteristic order
of the solutions
 within the same time interval  (here $10^4$)
persists. A variation of the noise
in steps of $10^{-4}$ for intervals within $[0.061,0.069]$ showed the same type
of ordered solutions.

Again, all runs for different noise intensities started from the very same
randomly chosen initial configuration of oscillator phases.
Also here a slight change in the initial conditions may lead to different
patterns as a function of the same increasing noise intensity.
We observed similar pronounced non-monotonic behavior for about
ten out of hundred different randomly chosen initial conditions.

\begin{figure*}
\begin{center}
\includegraphics[width=160mm]{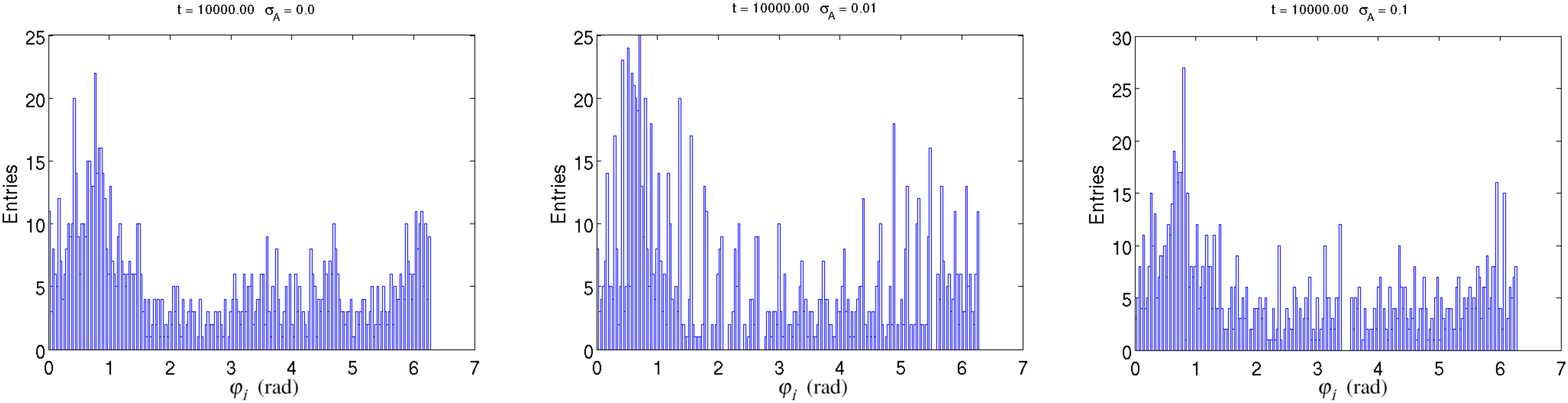}
\end{center}
\caption{Histogram of active rotator phases on a $32 \times 32$ lattice, 
at the time instant  $10^4$ time units after start of the integration, 
for parameter values $\omega = 0.7$,  $b=1$, and $\kappa = -2$, and additive 
noise, for three values of $\sigma_A$.}
\label{fig.6}
\end{figure*}

We have also identified {\it order-by-disorder} in larger systems
like $32\times 32$-lattices. Here we used histograms of the distribution
of phases over the unit circle, which result from a collection of phase
values  at a fixed time instant (here $10^4$ time units
from the start of integration.)
For an intermediate value of the noise intensity ($\sigma_A=0.01$),
the histogram has a more pronounced peak structure
than for vanishing noise or for a strength of $\sigma_A=0.1$,
see Fig.~\ref{fig.6}.
For such systems we expect a multitude of attractors
that is hard to resolve in its variety.

\begin{figure*}
\begin{center}
\includegraphics[width=140mm]{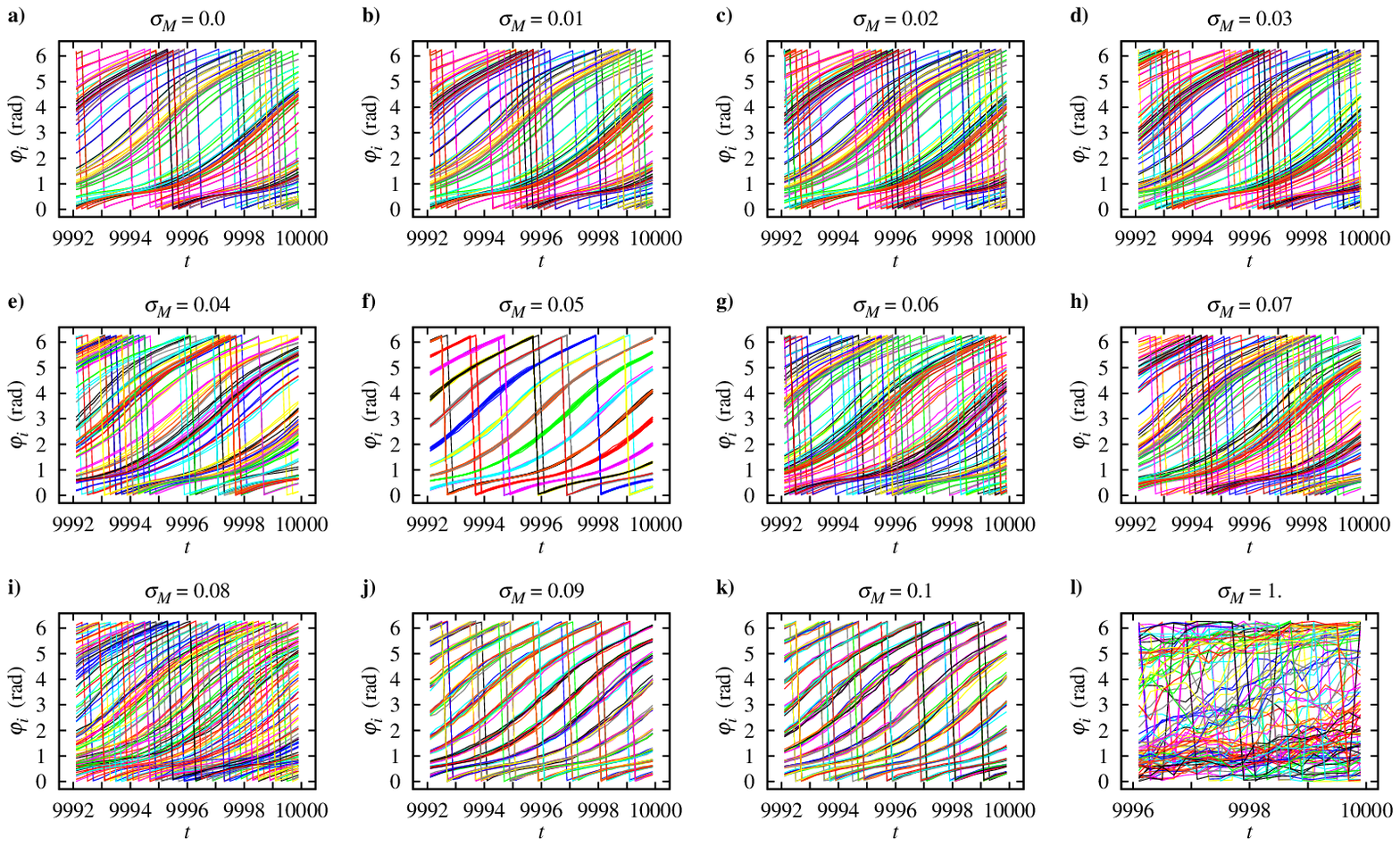}
\end{center}
\caption{Order-by-disorder on a $10 \times 10$ lattice of active rotators, for parameter values
$\omega = 0.7$,  $b=1$, and $\kappa = -2$, and multiplicative noise.}
\label{fig.7}
\end{figure*}

\emph{Order-by-disorder for multiplicative noise.} Similar sequences of
alternating synchronization patterns as in Fig.~\ref{fig.3} are obtained, when
the additive noise is replaced by multiplicative noise and $\sigma_M$ is varied
between $0.01$ and $0.1$, with other parameters fixed at the same values. Here
we again observe the effect of order-by-disorder for roughly $10\%$ of the
different initial conditions; an example is shown for active rotators in Fig.~\ref{fig.7}.

\begin{figure*}
\begin{center}
\includegraphics[width=140mm]{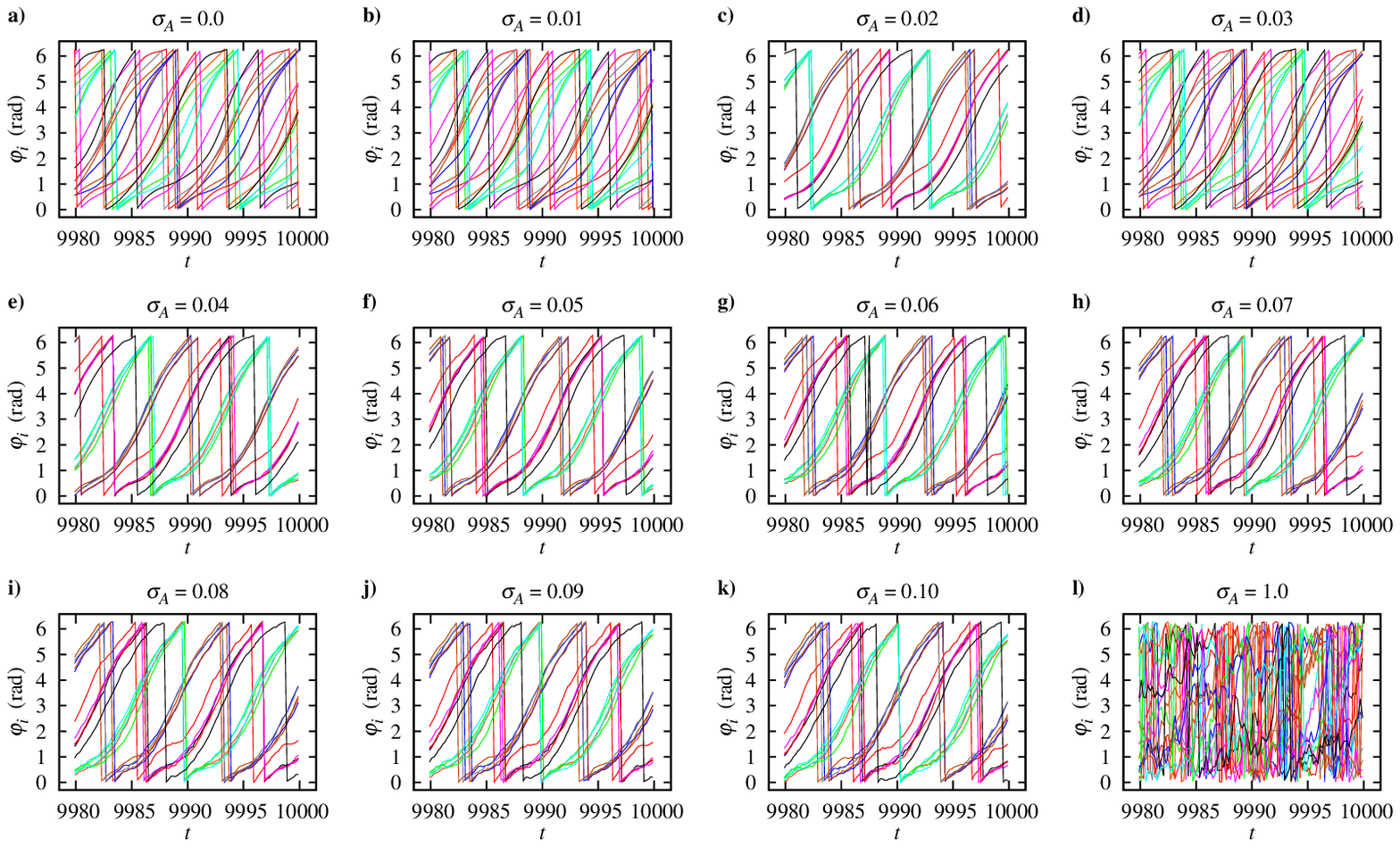}
\end{center}
\caption{Order-by-disorder on a $4 \times 4$ lattice of active rotators with free boundary conditions, for parameter values
$\omega = 0.7$,  $b=1$, and $\kappa = -2$. Same parameters as in Fig.~\ref{fig.3}, but with free boundary
conditions.}
\label{fig.8}
\end{figure*}

\emph{Order-by-disorder for non-periodic boundary conditions.}
According to our numerical data, the discussed effect seems to be unrelated
to the kind of boundary conditions on the lattice. As shown in Fig.~\ref{fig.8}, the tendency
to formation of long-living clusters at small and moderate values of
noise intensity persists for the lattice with free boundaries as well.

\emph{Order-by-disorder for Kuramoto oscillators.} Kuramoto oscillators on the
hexagonal lattice display similar sequences of states as in Fig.~\ref{fig.3},
confirming that the phenomenon is not caused by the excitability of the active
rotators, but by the structure of the underlying potential landscape, which
itself depends on the lattice topology and the sign-assignments of couplings.
In Table~\ref{table1} we illustrate why an alternative representation in terms of higher Kuramoto order parameters (here $\rho_1$, $\rho_2$, $\rho_4$, and $\rho_6$) fails to reflect the order of states in full detail.
\begin{table}
\centering
\begin{tabular}{|c|c|c|c|c|c|}
\hline
$\sigma_{A}$  & panel&  $\rho_{1}$  &  $\rho_{2}$  &  $\rho_{4}$  &  $\rho_{6}$  \\
\hline
\hline
0.00 &a &  0.001  &  0.102  &  0.060  &  0.149  \\
0.01 & b &  0.002  &  0.279  &  0.300  &  0.339  \\
0.02 & c &  0.003  &  0.868  &  0.514  &  0.104  \\
0.03 & d&  0.005  &  0.489  &  0.803  &  0.414  \\
0.04 & e&  0.006  &  0.420  &  0.503  &  0.366  \\
0.05 & f&  0.007  &  0.372  &  0.334  &  0.393  \\
0.06 & g&  0.009  &  0.838  &  0.461  &  0.235  \\
0.07 & h&  0.009  &  0.280  &  0.302  &  0.258  \\
0.08 & i&  0.011  &  0.767  &  0.303  &  0.227  \\
0.09 &j&  0.012  &  0.644  &  0.308  &  0.417  \\
0.10 & k&  0.013  &  0.787  &  0.346  &  0.224  \\
1.00 & l&  0.122  &  0.256  &  0.223  &  0.222  \\
\hline
\end{tabular}

\caption{Order parameter $\rho_1$, $\rho_2$, $\rho_4$, $\rho_6$
for the different noise intensities and their phase evolutions 
as displayed in Fig.~\ref{fig.9}.}
\label{table1}
\end{table}

\begin{figure*}
\begin{center}
\includegraphics[width=140mm]{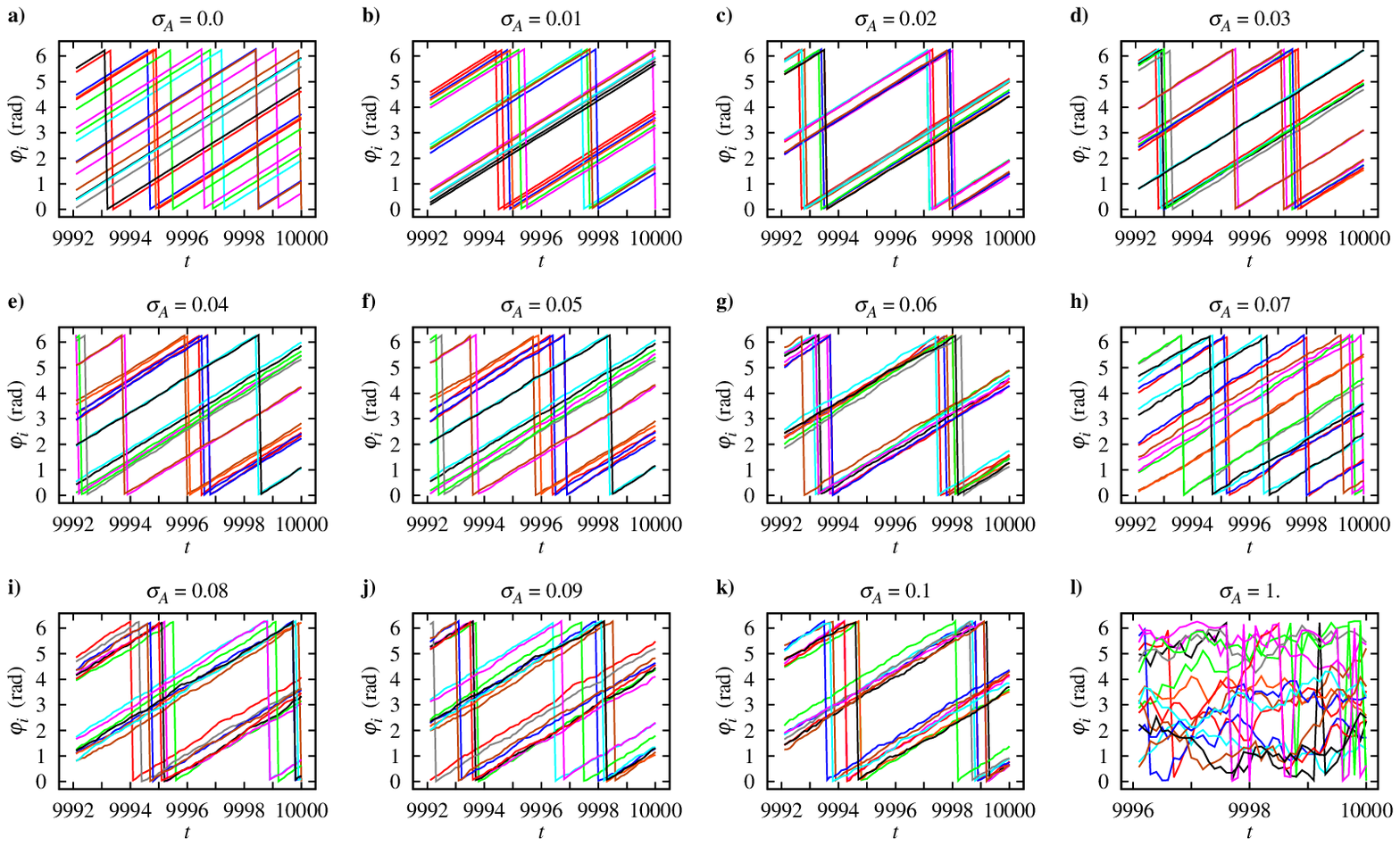}
\end{center}

\caption{
Order-by-disorder for Kuramoto oscillators on a $4 \times 4$ lattice, for parameter values
$\omega = 0.7$,  $b=0$, and $\kappa = -2$. Same parameters as in Fig.~\ref{fig.3} except for $b=0$.}
\label{fig.9}
\end{figure*}

For example, in the noisy disordered state ($\sigma_A=1.0$) they show
larger values than for the initial state of vanishing noise intensity
in panel (a) of Fig.~\ref{fig.9}.
While otherwise $\rho_2$ adequately reflects the change in the order
as observed in Fig.~\ref{fig.9}, $\rho_4$ fails for $\sigma_{A}=0.09$,
where the order is less than for $\sigma_{A}=0.08$ in contrast to what
the order parameter suggests.

\section{Hierarchies in the potential barriers}
\label{sec:hierarchies-in-the-potential}
We interpret the occurrence of
more ordered and less ordered cluster partitions for monotonically increasing
noise strength as an indication of a hierarchy in the potential barriers. In our
system the role of a potential is played by the integral over $\varphi$ of the
equations (\ref{eq_model}), in terms of which the equations obey a gradient dynamics,
$d\varphi_{i}/dt=-\nabla_i V$, with $V$ given by
\begin{equation}
V=-\omega\sum_{i}\varphi_{i}-b\sum_i\cos\varphi_{i}-\frac{\kappa}{2\mathcal{N}\,}
\sum_{i,j}A_{ij}\cos(\varphi_{j}-\varphi_{i}).
\label{eq:potential_function}
\end{equation}

The first term in (\ref{eq:potential_function}) is linear with respect to the
phases $\varphi_{i}$ and is responsible for the unbounded average drift whose
slope $\omega$ is the same for all patterns. Two last terms, taken together,
form the oscillatory part of the potential, $V_{\rm osc}$. We evaluate $V_{\rm
osc}$ for the 4-cluster solution, the 6-cluster solution, and the 16-cluster
solution. The mean values over one period are  $V_{\rm osc}^{(4)} = -7.76$,
$V_{\rm osc}^{(6)}= -7.65$, and $V_{\rm osc}^{(16)}= -7.51$, respectively: the
higher the order, the lower the corresponding potential. However, we possess no
detailed knowledge about  the landscape in between, in particular about the
height of the ridges. Sensitive response to different noise intensities
indicates that the landscape is quite rough, owing to the implemented high
degree of frustration. Numerical evidence suggests that for a sizeable portion
of initial conditions, deterministic paths to their eventual attractors pass
close to one or several ridges, which makes them sensitive to the action of
noise and introduces uncertainty in the destination.

In general, on the very large timescale,  the time evolution of $\varphi_i$
under the action of noise is a walk over the entire landscape. However, for the
small and moderate noise intensities the residence times in vicinities of the
deep minima or rifts are quite large (of the order of several hundred or
thousand time units), whereas the shallow valleys are traversed relatively fast.
Below we restrict ourselves to the moderately long epochs of evolution; hence,
for shortness, these intermediate asymptotics near which the system spends  long
intervals of time, are referred to as ``attractors''. It is these states that
are displayed in Fig.~\ref{fig.3}. Furthermore, as indicated before, it is not
the number of different patterns, characterized as $p_4,p_6,p_{16}$, or $d$,
which defines the number of different attractors, since characterizations in
terms of a pattern do not uniquely characterize a state due to the high
degeneracy in assigning a pattern to the grid. Therefore, what we call ``new
attractors" below may share the same pattern as encountered before.

\begin{figure*}
\centering
\includegraphics[width=160mm]{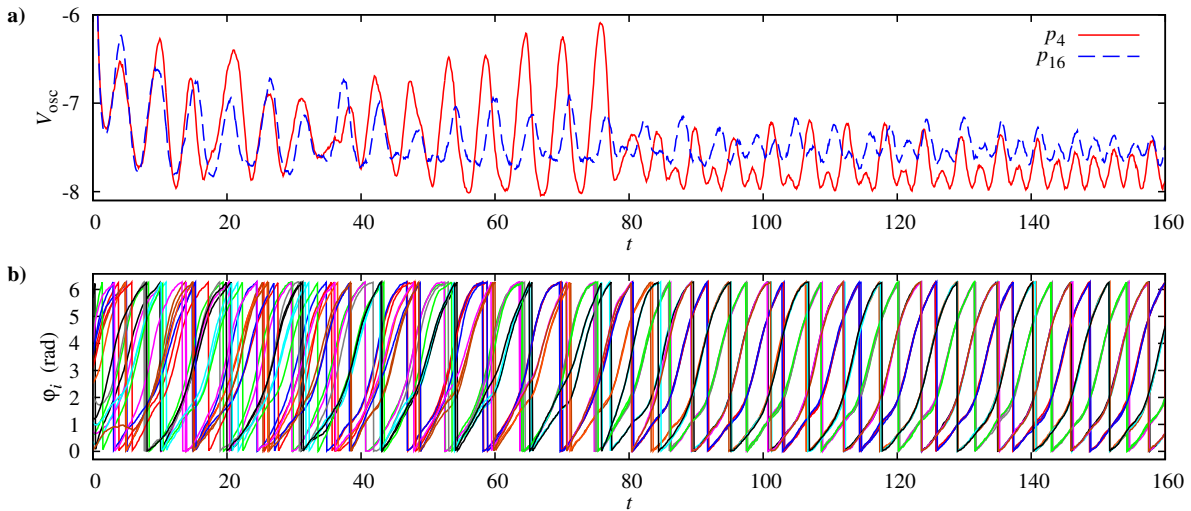}

\caption{\label{fig.10} a) Oscillatory component of the potential $V_{\rm
osc}(t)$ as function of time for $\omega = 0.7$,  $b=1$, $\kappa = -2$ and
$\sigma_M$=0, for two different realizations of additive noise at $\sigma_A=0.04$.
Starting from the same location with $V_{\rm osc}(0)=-0.248$ (not displayed),
two trajectories eventually go apart and end up at different attracting
patterns. The dashed line tracks the deterministic solution which leads to the
$p_{16}$-pattern. The solid line eventually leaves this path and goes to a
four-cluster state $p_4$. b) Individual phases $\varphi_i(t)$ for the solid
line from the top panel. Note the proximity to the three-cluster state in the
interval $45<t<75$. }
\end{figure*}

Consider an ensemble of stochastic trajectories which start from the same
initial location in the high-dimensional landscape, that is, the ensemble is
formed by different realizations of noise. Without noise, the gradient
dynamics drives the phases along the gradient of the potential until the first
local minimum or rift is met, where  the system gets stuck, be it in a fixed
point or in a limit cycle. Under sufficiently weak noise, the bulk of
the trajectories of the ensemble remains close to the deterministic orbit and
ends up at the same attracting set. Under somewhat stronger noise, the
ensemble splits: part of the trajectories crosses the nearby ridge(s) and goes
to different attractor(s), in contrast to the rest which follows the
deterministic trajectory. This is what we have seen to occur in ten
realizations. Fig.~\ref{fig.10}~(a) displays the temporal evolution of the
oscillatory part of the potential $V_{\rm osc}(t)$ for two trajectories which
start from the same initial position under the same parameter values and
correspond to two different realizations of additive noise at  $\sigma_A=0.04$.
The dashed line shows the trajectory which largely follows the deterministic
solution  and ends up at a $p_{16}$ pattern (the attracting state for these
initial conditions in the absence of noise). The trajectory which is shown by
the solid line, initially tracks the deterministic solution as well; however,
after a certain time and a short epoch of strong oscillations, it appears to
cross the ridge and land upon the deeper lying attractor of a $p_4$ solution. A
closer look at the individual phases (Fig.~\ref{fig.10} (b)) shows that
transitional oscillations are close to the three-clustered state: apparently
this is the unstable periodic solution, located on the ridge which separates the
basins of  $p_{16}$ and $p_4$. The trajectory approaches this state along its
stable manifold, spends some time (roughly the interval $45<t<75$) in its
neighborhood, displaying larger amplitudes of $V_{osc}$, and leaves it along the
unstable manifold. Very similar type of crossover behavior was found for
$\sigma_A=0.02$ and $0.03$ for different noise realizations, also leading to
$p_4$ or $p_{16}$ patterns in the end.

Starting with yet a higher noise level, the overwhelming majority of
trajectories cannot  resolve the former basin of attraction any longer, as its
shape gets buried under the noisy background; instead, the stronger noise
enables the trajectories to explore the phase space in more remote regions from
the starting point. It then depends on the depth of the valley whether the new
attractor is able to keep the trajectory in its vicinity for a while and let the
system settle inside the rift. Were the basin  as shallow as the former one, the
structure of the rift could not be recognized, and the transient passage of such
a basin would not be identified as long-living metastable state in our
simulations.

Now, as a matter of fact, our system finds repeatedly new attractors when the
noise is increased. This may be either due to the presence of several ridges
already in the vicinity of the starting point; it is then a random event which
attractor is chosen under a new realization of noise, once its depth is
sufficiently large. Or the new attractors are discovered when the noise drives
the system a longer path through phase space, as long as an attractor in a more
remote deep valley stops the walk for a transient time.  Wherever the new
attractors are located, with increasing noise they have to be increasingly deep
to become observable. Naturally, trajectories visiting  basins of attractions
between the higher ridges  will be less localized for stronger noise. This
feature explains the need for adapting the size of the tolerance interval within
which two phases are identified, see Fig.~\ref{fig.3}~(d) and~(f).

For sufficiently high noise intensities,
the whole potential structure $V$ is buried under the noise, and the phase
trajectory performs a random walk, now driven by an effectively random
potential, without correlations between individual phases, as it was seen in
Fig.~\ref{fig.3}~(h). \vskip5pt It should be noticed that Fig.~\ref{fig.10}(b)
shows only the first 160 time units of the quasi-stationary state to illustrate
the dispersion of the trajectories after about $45$ time units. The state
remains time-dependent afterwards, as the system keeps jumping from one
attractor to another in the presence of noise. The histogram of escape times
from one type of ordered state to another (say from any of the $p_4$
to any of the $p_{16}$-states) shows a multi-peak structure.
More details will be presented in a forthcoming paper.

\subsection*{Analogy to stochastic resonance}
\begin{figure*}
\begin{center}
\includegraphics[width=120mm]{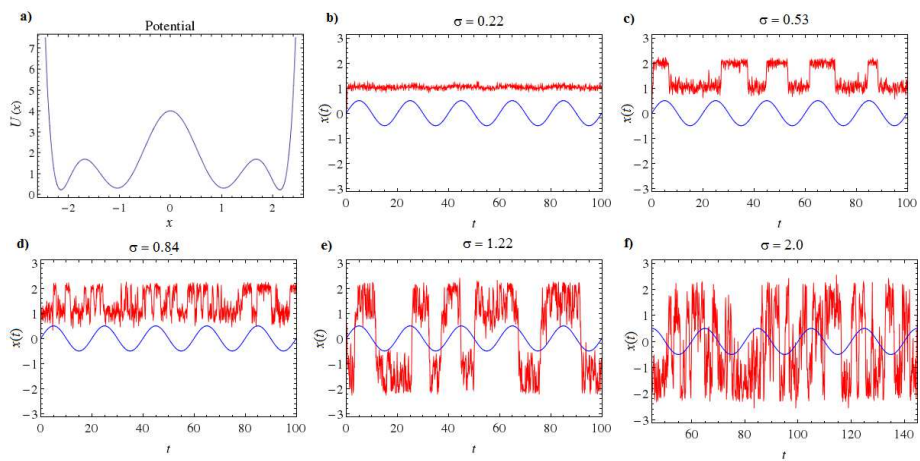}
\end{center}
\caption{Stochastic resonance for a potential (a) with two barrier heights:
Panels (b) - (f) show the value of the force (in blue) and the response of the
system (in red) as functions of time, for different noise intensities
$\sigma = 0.22, 0.53, 0.84, 1.22,$ and $2.0$.
For further explanations see the text.}
\label{fig.11}
\end{figure*}

As an argument in favor of our conjecture  of  a hierarchical landscape
with potential barriers of various depths,
we demonstrate analogous behavior for stochastic
resonance. To this aim we choose a one-dimensional potential $U(x)$ with only
two levels of barriers: $U(x)=\sum_{i=1}^4\alpha_i x^{2i}$, as shown in
Fig.~\ref{fig.11}~(a). Unlike most studies of stochastic resonance, we are
interested not in temporal (resonance-like) aspects of the dynamics, but in the
localization of solutions in different basins.

\begin{figure*}
\begin{center}
\includegraphics[width=140mm]{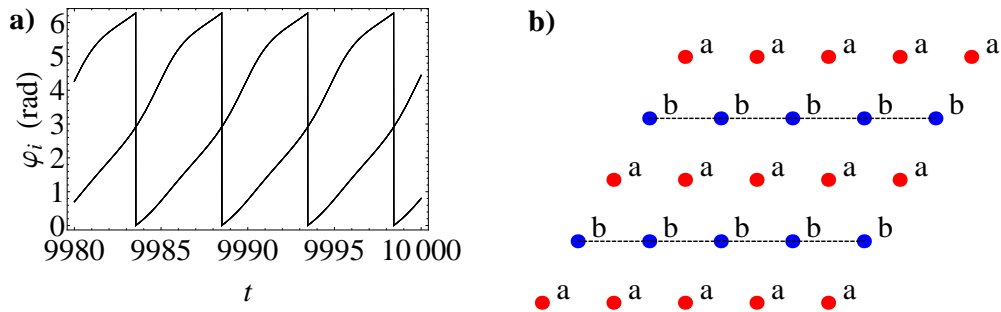}
\end{center}
\caption{Two-cluster solution of active rotators on a $4\times4$ lattice without frustration.
Oscillators in cluster $a$ and $b$ share the same phase, respectively. The
solution is representative  for our results for $\omega = 0.7$,  $b=1$, and
$\kappa = -2$ and $10^6$ randomly chosen initial conditions. }
\label{fig.12}
\end{figure*}

Consider the motion of a particle, described by the stochastic differential
equation \begin{equation} \frac{dx}{dt}\;=\;-\frac{dU}{dx}+ A \sin{\omega t}+
\xi(t), \end{equation} where $\xi(t)$ is Gaussian white noise with intensity
$\sigma$: $\langle \xi(t)\rangle=0$, $\langle \xi(t)\xi(t^\prime)\rangle = 2
\sigma \delta(t-t^\prime)$. An external subthreshold  periodic perturbation
$A\sin{\omega t}$ does not allow the particle to leave any of the four local
minima in the absence of noise, but combined with noise it triggers the
switching between different minima. In Fig.~\ref{fig.11}~(b)-(f) the response of
the system to increasing noise intensity $\sigma$ is shown. For low intensity
($\sigma=0.22$), we observe very small random oscillations about one of the
local minima, here about $x=1$ (Fig.~\ref{fig.11}~(b), red line). Regular
oscillations in Figs.~\ref{fig.11}~(b)-(f) show the periodic external force (blue
lines). For an optimal noise strength ($\sigma \simeq 0.53$), the particle is
enabled to cross the lower barrier, so it jumps in resonance with the external
frequency $\omega$ between  $x=1$ and $x=2$ (or, depending on the initial
conditions, $x=-1$ and $x=-2$) (Fig.~\ref{fig.11}~(c)). For an intermediate
larger noise intensity the motion between $x=1$ and $x=2$ is irregular
(Fig.~\ref{fig.11}~(d)). For crossing the second barrier, there is the second
``optimal" noise strength $\sigma \simeq 1.22$ for which the particle jumps in
resonance with the external frequency between locations in the interval $[1,2]$
and $[-2,-1]$ (Fig.~\ref{fig.11}~(e)). For even larger noise the particle does no
longer see the underlying shape of the potential, but moves irregularly between
the outer walls (Fig.~\ref{fig.11}~(f)). We see that under higher noise intensity
the localization occurs on levels separated by higher barriers, and the particle
is less localized in the vicinity of the minima. This reduced localization is
akin to our need for adapting the tolerance interval when two rotator phases are
identified within an uncertainty of $\pm\Delta\varphi$.

\section{Disorder in the coupling signs, but no frustration}
\label{sec:disorder-in-coupling}
Without frustration, but with disorder in the coupling signs according to the choice as
in Fig.~\ref{fig.1}~(b), we have neither seen a signature for multistability,
nor for the order-by-disorder phenomenon in oscillators. On a $4\times4$-lattice
with otherwise the same choice of parameters we started from $10^6$ randomly
chosen initial conditions, and from $10^4$ for a $10\times10$-lattice. The only
stationary state we have found were 2-cluster solutions with oscillators in
cluster $A$ aligned along a horizontal line and alternating with oscillators in
cluster $B$ along the succeeding horizontal line, where the phase difference
between clusters $A$ and $B$ fluctuates about $\pi$ (see Fig.~\ref{fig.12}).

\section{Summary and conclusions}
\label{sec:conclusions}
It is well known that the role of noise in
nonlinear systems can be quite versatile \cite{lutz}: in particular, it can
increase the order in a disordered system. Here we made use of a similar
mechanism that is known from spin systems with an analogous role played by the temperature there and noise here: we have chosen the topology and the
couplings in a way as to induce frustrated bonds. Frustration leads to a
considerable increase in the number of attracting states. Since the units are
oscillatory, coexisting states are no longer restricted to fixed points as in
spin systems, but can also be different patterns of phase-locked motion. The
energy landscape is not necessarily degenerate, but the system is multistable.
Similarly to spin systems, here the noise can temporarily increase
the order of states, a fact that we called order-by-disorder. Surprisingly,
however, as a function of monotonically increasing noise intensity we observe
not only a maximum of order at some optimal noise intensity, as one may have
expected from analogous results on stochastic resonance \cite{stoch}, coherence
resonance \cite{coh}, or system size resonance \cite{sys}. Instead we record an
alternating sequence of higher and lower order among the oscillator phases, so
that the variation of the noise intensity appears to ``scan'' a rough potential
landscape with a hierarchy of potential barriers. To turn it into a practical
device for scanning similar kinds of potential, the high sensitivity to the
initial conditions should be reduced by adding appropriate control terms to the
dynamics. Interestingly, the quasi-stationary state in the presence of noise
shows a variety of time scales, realized in the escape times from one to another
``attractor" (metastable state), which we shall further explore in a forthcoming
paper.

\acknowledgments
Two of us (F.I. and D.L.) would like to thank
the Deutsche Forschungsgemeinschaft (DFG, contract ME-1332/19-1) for financial
support. M.A.Z. was supported by the DFG Research Center MATHEON (Project D21).



\appendix


\section{}

\subsection{Multistability for plane waves in Kuramoto oscillators}
In the corotating
reference frame the existence of steady states for Kuramoto oscillators is
independent on the coupling strength, while their stability depends on the
coupling sign. Here we formulate sufficient stability conditions for a certain
class of fixed points,  and for the coexistence of multi-stable solutions.

\begin{figure*}
\centering
\includegraphics[width=140mm]{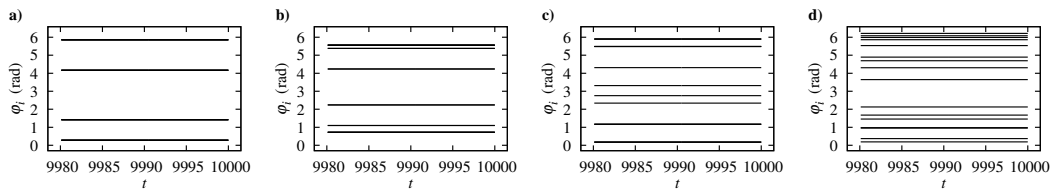}

\caption{Coexisting fixed point solutions with 4 clusters (a), 6 clusters (b), 8 clusters
(c), and 16 clusters (d) obtained for parameter values $\omega = 0$, $b = 1$,
and $\kappa = -2$. }
\label{fig.13}
\end{figure*}

On the hexagonal $M\times L$ lattice, the equation of motion for each oscillator
includes 6 terms.  We restrict ourselves to the plane waves with the pattern (\ref{eq_plane_wave}):
$\varphi_{m,n}=2\pi\,(k_1m/M+k_2n/L),\;m=1,\ldots\, M;\, n=1,\ldots,L$;
there, the terms cancel pairwise, hence every solution
of this form is an equilibrium point. Let the coupling strength $\kappa$ be negative.

In the symmetric Jacobian matrix for an equilibrium of this kind, each row includes
7 non-zero elements. The non-zero off-diagonal elements are, respectively,
two elements $-\cos 2\pi k_1/M$, two elements  $-\cos2\pi k_2/L$
and two elements $-\cos 2\pi (k_1/M+k_2/L)$. The diagonal elements equal
$\cos 2\pi k_1/M+\cos2\pi k_2/L+\cos 2\pi (k_1/M+k_2/L)$.
Since the sum in every row vanishes, the Jacobian always possesses the zero eigenvalue
which corresponds to the translational invariance of the equations of motion
(\ref{eq_kuramoto}).

If all off-diagonal elements are negative, that is,
\begin{equation}
\cos\frac{2\pi k_1}{M}<0,\;\cos\frac{2\pi k_2}{L}<0,\;
\cos 2\pi\left(\frac{k_1}{M}+\frac{k_2}{L}\right)<0,
\label{condition}
\end{equation}
then the Jacobian matrix has no positive eigenvalues. Indeed, by splitting the Jacobian
into the diagonal and the off-diagonal matrices we observe, as a consequence of the
Frobenius-Perron theorem, that none of the eigenvalues of the non-diagonal part
exceeds in the absolute value the (negative) diagonal element. This implies
stability for the considered plane wave solution.

Note that this condition is not necessary: it does not include e.g.
the patterns of spherical waves, which are numerically observed as stable
solutions as well.

Next we use the same framework to demonstrate coexistence
on sufficiently large lattices
of plane waves (\ref{eq_plane_wave}) which differ by their number of clusters.
For the case of the square lattice with $M=L, k_1=k_2$, condition
(\ref{condition}) is reduced to $L/4<k<3L/8$.
For $L\geq 18$ this interval contains at least two integers.
Taking an even $L\geq$18 and two consecutive integers $q$ and $q$+1
from the stability interval,
we observe that the first one corresponds to the pattern with $L/\gcd(L,\, q)$ clusters,
where $\gcd(L,\, q)$ is the greatest common divisor of $L$ and $q$,
whereas the second one corresponds to  $L/\gcd(L,\, q+1)$ clusters.
Since two consecutive integers have no nontrivial common divisors,
whereas a divisor 2 is shared between $L$ and either $q$ or $q$+1,
the numbers of clusters in two stable steady patterns cannot coincide.

\subsection{Stability of the ``synchronous" fixed point for active rotators}
Here we discuss the synchronous steady solution for Eq.(\ref{eq_model}).
We set, without restrictions of generality, $b$=1
and work in the parameter space of $\omega$ and $\kappa$.

We start from the non-coupled case $\kappa=0$. For
$\omega < 1$ there are $2^{N}$ fixed points,
defined by $\varphi_{i}=\varphi_{1,2}^{*}$, with
$\varphi_{1}^{*}=\arcsin \omega $ and $\varphi_{2}^{*}=\pi-\arcsin(\omega)$.
For linearizations of (\ref{eq_model}) at these points,
the number of positive eigenvalues of the respective Jacobian
equals the number of components $\varphi_{2}^{*}$.
Accordingly, out of $2^{N}$ steady states
$2^{N}$--2 are saddles, one is stable, and one is unstable.
For the stable equilibrium the values of each coordinate equals $\arcsin\,\omega$.

Now we switch on the coupling $\kappa$. Notably, the
location of this equilibrium in the phase space is independent of $\kappa$.
For the corresponding Jacobian $J$ we obtain
\begin{equation}
J_{ij}=\begin{cases} {\displaystyle -\sqrt{1 -
\omega^{2}}-\kappa}, & \qquad i=j\\
{\displaystyle \frac{\kappa}{\mathcal{N}}A_{ij}} &
\qquad i\not=j \;,\end{cases}
\end{equation}
so that
\begin{equation}
J=\frac{\kappa}{\mathcal{N}}A-(\sqrt{1 - \omega^{2}}+\kappa)I,
\end{equation}
where $\mathcal{N}$, as above, is the number of nearest neighbors on the lattice.
The eigenvalues of $J$ are
\begin{equation}
\lambda_{i}=\frac{\kappa}{\mathcal{N}}\lambda_{i}^{A}
-\kappa-\sqrt{1-\omega^{2}},
\label{eq:lam-i-neg}
\end{equation}
where $\lambda_{i}^{A}$ are the eigenvalues of the adjacency matrix.
For $\kappa$=0, all eigenvalues of $J$ are negative.
For $\kappa<0$, the first eigenvalue $\lambda_{i}$ that becomes positive corresponds
to the {\em minimal} eigenvalue of $A$, $\lambda_{\text{min}}^{A}$, provided
$\lambda_{\text{min}}^{A}<\mathcal{N}$.
The transition takes place at
\begin{equation}
\kappa_{c}=-\frac{\sqrt{1^2-\omega^{2}}}{1-\lambda_{\text{min}}^{A}/\mathcal{N}};
\label{eq:kc-neg}
\end{equation}
here the stable fixed point loses stability and becomes a saddle.
Notably, for a hexagonal lattice ($\mathcal{N}$=6) the minimal eigenvalue
is always degenerate.

Our numerical observations on the
hexagonal lattices suggest that the synchronous equilibrium has a
large basin of attraction for sufficiently small weak negative
couplings: this attractor was the only one which we found for $10^4$
different randomly chosen initial conditions on various lattice sizes
($3\times3$, $4\times4$, $5\times5$ and $\omega=0.7$) and $\kappa=-0.5$, $-0.7$,
$-0.9$. On the other hand, for values $\kappa<\kappa_c$, multistable solutions were easily found by starting from different initial conditions, solutions with clusters of phase locked oscillators that performed limit cycles. Therefore one may wonder whether multistable fixed point solutions exist as well. Such coexisting fixed point solutions can be identified for the case
of $\omega=0$, starting from initial
conditions, which are given by phase distributions that are known from Kuramoto
oscillators for ($b=0$) and otherwise the same choice of parameters, but then
turning on $b$. Since we can absorb $b\not=0$ in the time scale (setting $b=1$),
we checked the stability of these fixed point solutions under tuning $\kappa$ to
more negative values, keeping $\omega=0$. Using the Newton Raphson method, we
found coexisting fixed points for
$\kappa<\kappa_c=-1/(1-\lambda_{\text{min}}^A/6)$ which differ in the cluster partition of coinciding phases.
For the coexisting fixed-point solutions see Fig.~\ref{fig.13}.


\begin{thebibliography}{37}

\bibitem{26}
J. Villain, R. Bidaux, J.P. Carton, R. Conte,
J. Phys. (France) \textbf{41}, 1263 (1980)


\bibitem{29}
D. Bergman, J. Alicea, E. Gull, S. Trebst, L. Balents,
Nature Physics \textbf{3}, 487 (2007)


\bibitem{barnett}
R. Barnett, S. Powell, T. Gra{\ss}, M. Lewenstein, S. Das Sharma,
Phys. Rev. A \textbf{85}, 023615 (2012)


\bibitem{reimers}
J.N. Reimers, A.J. Berlinsky,
Phys. Rev. B \textbf{48}, 9539 (1993)


\bibitem{chubukov}
A. Chubukov, Phys. Rev. Lett. \textbf{69}, 832 (1992)


\bibitem{27}
R. Moessner, J.T. Chalker, Phys. Rev. B \textbf{58}, 12049 (1998)


\bibitem{28}
C.L. Henley, Phys. Rev. Lett. \textbf{ 62}, 2056 (1989)


\bibitem{30}
A.M. Turner, R. Barnett, E. Demler, A. Vishwanath,
Phys. Rev. Lett. \textbf{98}, 190404 (2007)


\bibitem{15}
H. Sakaguchi, S. Shinomoto, Y. Kuramoto, 
Prog. Theor. Phys. \textbf{79}, 600 (1988)


\bibitem{183}
C.J. Tessone, D.H. Zanette, R. Toral,
Eur. Phys. J. B \textbf{62}, 319 (2008)


\bibitem{zanette}
D.H. Zanette, Europhys. Lett. \textbf{72}, 190 (2005)


\bibitem{hong}
H. Hong, S.H. Strogatz,
Phys. Rev. E \textbf{84}, 046202 (2011)


\bibitem{daido}
H. Daido, Phys. Rev. Lett. \textbf{68}, 1073 (1992)


\bibitem{pablo}
P. Kaluza, H. Meyer-Ortmanns,
Chaos \textbf{20}, 043111 (2010)


\bibitem{Hong_Park}
H. Hong, H. Park, M.Y. Choi,
Phys. Rev. E \textbf{72}, 036217 (2005)

\bibitem{lutz}
B. Lindner, J. Garcia Ojalvo, A. Neiman, L. Schimansky-Geier,
Physics Reports \textbf{392}, 321 (2004)


\bibitem{stoch}
L. Gammaitoni, P. H\"anggi, P. Jung, F. Marchesoni,
Rev. Mod. Phys. \textbf{70}, 223 (1998)


\bibitem{coh}
A.S. Pikovsky, J. Kurths,
Phys.Rev.Lett. \textbf{78}, 775 (1997)


\bibitem{sys}
R. Toral, C.R. Mirasso, J.D. Gunton,
Europhys. Lett. \textbf{61}, 162 (2003)


\end{thebibliography}
\end{document}